\documentclass[aps,prb,reprint,10pt]{revtex4-1}

\usepackage{amsmath}
\usepackage{amssymb}
\usepackage{amsthm}
\usepackage[pdftex]{graphicx}

\newcommand{\bra}[1]{\langle {#1} \vert}
\newcommand{\ket}[1]{\vert {#1} \rangle}

\newcommand{\braket}[2]{\langle {#1} \vert {#2} \rangle}

\begin{document}

\title{Measurement- and comparison-based sizes of Schr\"{o}dinger cat states of light}

\author{T.J. Volkoff}
\author{K.B. Whaley}
\affiliation{Berkeley Quantum Information and Computation Center and Dept. of Chemistry, UC Berkeley}

%\date{\today}

\begin{abstract}
We extend several measurement-based definitions of effective ``cat-size" to coherent state superpositions with branches composed of either single coherent states or tensor products of coherent states. These effective cat-size measures depend on determining the maximal quantum distinguishability of certain states associated with the superposition state: \textit{e.g.}, in one measure, the maximal distinguishability of the branches of the superposition is considered as in quantum binary decision theory; in another measure, the maximal distinguishability of the initial superposition and its image after a one-parameter evolution generated by a local Hermitian operator is of interest. The cat-size scaling with the number of modes and mode intensity (\textit{i.e}. photon number) is compared to the scaling derived directly from the Wigner function of the superposition and to that estimated experimentally from decoherence. We also apply earlier comparison-based methods for determining macroscopic superposition size that require a reference GHZ state. The case of a hierarchical Schr\"{o}dinger cat state with branches composed of smaller superpositions is also analyzed from a measurement-based perspective.
\end{abstract}

%\pacs{06.20.-f 03.65.Ud 03.65.Yz}
%\keywords{}

\maketitle

\section{\label{sec:intro}Introduction}
The notion of a Schr\"{o}dinger cat state arises from extrapolating the quantum superposition principle for microscopic (\textit{e.g.}, small de Broglie wavelength) systems  to classically distinguishable macroscopic states. In physical systems comprised of a large ($N\gg 1$) but finite number of identical, distinguishable degrees of freedom, i.e. the total Hilbert space of the system $\mathcal{H}_{\text{sys}}=\mathcal{H}^{\otimes N}$ is  the $N$-th tensor product of the single-particle Hilbert space $\mathcal{H}$, the $\ket{\text{GHZ}_{N}}$ state comprised of orthogonal single particle states $\ket{\phi_{1}}$, $\ket{\phi_{2}} \in\mathcal{H}$ are the prototypical examples of cat states:
\begin{equation}\label{eqn:ghz} \ket{\text{GHZ}_{N}} := {1\over \sqrt{2}}\left( \ket{\phi_{1}}^{\otimes N} + \ket{\phi_{2}}^{\otimes N} \right) \end{equation}

Given a measure of superposition size with definite value (say, $N$) for $\ket{\text{GHZ}_{N}}$, it is natural to determine the cat sizes of generalized $\ket{\text{GHZ}_{N}(\epsilon)}$ states, macroscopic superpositions of the same form as Eq.(\ref{eqn:ghz}) except with $\vert \braket{\phi_{1}}{\phi_{2}} \vert^{2} = 1-\epsilon^{2}$, $\epsilon < 1$, and appropriate normalization. D\"{u}r, Simon, and Cirac  \cite{cirac} have shown that an effective cat size $N\epsilon^{2}$ can be attributed to these states by comparing the actions of three maps on the $\ket{\text{GHZ}_{N}(\epsilon)}$ state to their actions on an ideal $\ket{\text{GHZ}_{N}}$ state: 1) decoherence to the macroscopic mixed state, 2) distillation to a $\ket{\text{GHZ}_{n(N)}}$ state as in Eq.(\ref{eqn:ghz}) but with $n(N)<N$, and 3) particle loss according to binomial statistics. The same effective cat size scaling has been derived by other authors by calculating the maximal quantum distinguishability of the branches of the superposition,\cite{whaleyjan} by calculating the maximal quantum Fisher information \cite{dur} per mode of the superposition, and by considering unitary evolution (generated by an appropriately chosen operator) of the superposition to an orthogonal state. \cite{bjork} In each of these cases the calculation of superposition size depends on locating a positive, self-adjoint operator on the tensor product space which is optimal for the given cat-size definition and so we refer to these cat sizes as being defined in a ``measurement-based" way. Yet another measurement-based measure has been introduced which involves associating to each composition of $d$ single-particle operators, a probability $p(d)$ that this composition applied to one branch of a cat state produces the other branch. \cite{marquardt} These cat size measures contrast with the ``comparison-based" sizes of Ref.[\onlinecite{dur}], which require a reference superposition state.

In this paper, we show that the methods mentioned above can be extended to describe cat states with branches composed of photonic coherent states or superpositions of tensor products thereof (the latter are usually called entangled coherent states) and that a subset of the resulting cat sizes are consistent with empirical notions of cat size based on the quasiprobability distributions and decoherence of these states.\cite{jeong, harochebook} The semiclassical macroscopic states which are superposed to create these cats are parameterized by the amplitudes and phases of one or more classical electromagnetic modes; the Hilbert space of each mode is that of a quantum harmonic oscillator $\ell^{2}(\mathbb{C})$. \cite{reed} These intriguing superpositions have been generated experimentally using nonlinear optical Kerr media \cite{schoelkopf} or nonresonant coupling between Rydberg atoms and a high-$Q$ optical cavity, \cite{harochedavidovich} and their properties, including their quasi-probability distributions and photon number distributions, have been studied theoretically. \cite{sanders, enkhirota, vanenk2} Although notions of intrinsic size of coherent state cats based directly on static properties of their quasiprobability distributions in phase space or their evolution under decoherence have been considered, they have not been assessed in terms of the existing comparison-based or measurement-based approaches. In order to apply to coherent state cats, these comparison-based and measurement-based approaches must take into account subtle properties of coherent states, including indefinite particle (photon) number, the lack of natural subsystems of particles, and nonorthogonality of the coherent state basis of $\ell^{2}(\mathbb{C})$.

Previous authors have suggested that any notion of superposition macroscopicity (cat size) must be predicated on a choice of quantum property exhibited by the superposition which either is not exhibited by its branches individually \cite{dur, bjork} or which is otherwise important for the state to be considered macroscopic.\cite{marquardt} The quantum property is not necessarily unique; so one may form cat-size definitions of greater (weaker) strength by combining (separating) the conditions. For example, Leggett has combined the requirements of ``disconnectivity" of a superposition state with existence of a self-adjoint operator exhibiting an extensive difference in expectation value in the branches of the superposition to form his notion of macroscopic distinctness.\cite{leggett}  The measurement-based measures of cat size discussed in this paper are each predicated on a quantum property of the superposition. On the other hand, some quantum properties lead to recognition of superpositions which are clearly the most ``cat-like" and in these cases, comparison-based cat size measures lead to reasonable results for a given superposition. Finally, in a third class of superposition size measures, the quantum properties are intrinsic to the superposition state, \textit{e.g.}, a geometric property of a quasiprobability distribution associated to the state which is determined empirically or by an interference-based measure.

\section{\label{sec:results}Superposition size measures of photonic coherent states}

The Schr\"{o}dinger cat state we will be concerned with is an $N$-mode entangled coherent state:\cite{jeongkim, enkhirota}) \begin{equation}\label{eqn:entangledstate}
\ket{\Omega}={1\over \sqrt{2+ 2\exp(-2N\vert \alpha \vert^{2})}}\bigg( \ket{\alpha}^{\otimes N} + \ket{-\alpha}^{\otimes N} \bigg) .
\end{equation} This state is an analog in the Hilbert space $\ell^{2}(\mathbb{C})^{\otimes N}$ of the generalized spin states $\ket{\text{GHZ}_{N}(\epsilon)} \in (\mathbb{C}^{2})^{\otimes N}$ considered above, with $\epsilon = 1-\exp(-2\vert \alpha \vert^{2})$. The field amplitude $\alpha$ can be taken on the real line for convenience. The semiclassical branches of this cat state are composed of $N$ separate EM field cavities with amplitude $\pm \alpha$. It will be useful to note that $\ket{\Omega}$ with $N=2^{M}$ for $M$ a natural number can be generated by application of a sequence of beamsplitters and phase-shifters from, e.g., a state $\ket{\Omega '}$ composed of a product of a single-mode superposition of coherent states and auxiliary vacuum modes:\cite{nguyen} \begin{equation}\label{eqn:beam} \ket{\Omega '} \propto \left( \ket{\sqrt{N}\alpha}_{1} + \ket{-\sqrt{N}\alpha}_{1} \right)\otimes \bigotimes_{m=2}^{N}\ket{0}_{m} \end{equation} In fact, this equation is true for any positive integer $N$ (see Appendix \ref{sec:app}). This identity allows one to map the problem of determining cat size of a single-mode ($N=1$) coherent state superposition onto the same problem for entangled coherent states (and \textit{vice versa}) if one admits the following axiom: for any cat size measure, the cat size is invariant under appending auxiliary vacuum modes to the state and mixing these modes with the state. This axiom is important because cat size is clearly not invariant under arbitrary unitary operations in the product space (which can create a macroscopic superposition from, e.g. a product state), so this axiom identifies a set of unitaries under which the cat size is invariant.

\subsection{\label{sec:measurement}Measurement-based measures of cat size}
\subsubsection{\label{sec:korsbakken}Branch distinguishability measures}
The motivation behind the measurement-based superposition size measure of Ref.[\onlinecite{whaleyjan}] is the extension to general quantum measurements of binary decision theory.\cite{helstromone} Let $\ket{A}$, $\ket{B}\in (\mathbb{C}^{2})^{\otimes N}$ with $\Vert \ket{A} \Vert = \Vert \ket{B} \Vert$ and let $\rho_{A}$, $\rho_{B}$ be their respective pure states.

\underline{Definition 1}:  Given  $0<\delta <1/2$, the cat size $C_{\delta}(\ket{\psi})$ of $\ket{\psi} \propto \ket{A}+\ket{B}$ is \begin{equation} C_{\delta}(\ket{\psi}) = {N\over n_{\text{eff}}(\delta , \ket{\psi})} \end{equation} where \begin{equation} n_{\text{eff}}(\delta, \ket{\psi}) = \text{min}_{n}\lbrace {1\over 2} +{1\over 4}\Vert \rho_{A}^{(n)} - \rho_{B}^{(n)} \Vert_{1} > 1-\delta\rbrace \label{eqn:helstrom} \end{equation}

The $n$-reduced density matrix ($n$-RDM) $\rho^{(n)}_{A}$ is defined  by $\text{tr}_{N- n}\vert A \rangle \langle A \vert$ and $\Vert \cdot \Vert_{1}$ is the trace norm. The left side of the bracketed inequality in Eq.(\ref{eqn:helstrom}) is recognizable as the maximal success probability (over all $n$-mode positive operator-valued measurements) for distinguishing the $n$-RDMs $\rho_{A}^{(n)}$ and $\rho_{B}^{(n)}$. \cite{fuchs,helstromone} This measure is simplest to evaluate when the individual modes are distinguishable, i.e. every subsystem is individually addressable. A notable feature of using branch distinguishability in a measure of cat size is that the measurement which allows the binary decision success probability to be written as the trace norm of the difference between the (generally nonorthogonal) branches also effectively collapses the superposition defining the cat with high probability. This is demonstrated quantitatively for the even cat state $\ket{\psi_{+}} \propto \ket{\alpha} + \ket{-\alpha}$ in Appendix \ref{sec:appbranch}.

We now apply the measurement-based measure above to the $N$-mode entangled coherent state of Eq.(\ref{eqn:entangledstate}), which is in $\ell^{2}(\mathbb{C})^{\otimes N}$. To evaluate $C_{\delta}(\ket{\Omega})$, we write $\rho_{\pm \alpha} \equiv \vert \pm\alpha \rangle \langle \pm\alpha \vert$ and note that $\text{tr}_{N-n}(\rho_{\alpha}^{\otimes N} - \rho_{-\alpha}^{\otimes N}) = (\rho_{\alpha} - \rho_{-\alpha})^{\otimes n}$, which can be viewed as a product of measurements in the computational basis $\lbrace \ket{\alpha} , \ket{-\alpha}\rbrace $ when coherent states are used as qubits.\cite{milburn}  We do not have to change the Hilbert space from $(\mathbb{C}^{2})^{\otimes N}$ for which $C_{\delta}$ is defined to $\ell^{2}(\mathbb{C})^{\otimes N}$ because the relevant trace norm can be computed in an orthonormal basis $\lbrace \ket{e_{1}} = \ket{\alpha} , \ket{e_{2}} \propto \ket{-\alpha} - \exp(-2\vert \alpha \vert^{2})\ket{\alpha} \rbrace$ for a 2-D subspace of $\ell^{2}(\mathbb{C})$ which contains e.g. $\ket{\alpha}$, $\ket{-\alpha}$ and superpositions thereof, in particular the eigenvectors of $\rho_{\alpha} - \rho_{-\alpha}$. The resulting precision-dependent effective size and cat size are: \begin{eqnarray} n_{ \text{eff}}(\delta , \ket{\Omega}) &=& \Big\lceil {\log (4\delta - 4\delta^{2}) \over -4\vert \alpha \vert^{2}} \Big\rceil \nonumber \\ C_{\delta}(\ket{\Omega}) &\approx& {-4N\vert \alpha \vert^{2}       \over \log (4\delta - 4\delta^{2})  } \label{eqn:cat}\end{eqnarray}
 with the former quantity interpreted as the minimal number of modes which must be measured in order to successfully distinguish the branches of $\ket{\Omega}$ with probability $1-\delta$. The definition of the cat size through the partial trace over modes in the branches of the superposition requires $1\le  n_{ \text{eff}}(\delta , \ket{\Omega}) \le N$ to be satisfied. In turn, in order for this inequality to hold, it is required that the precision $\delta$ satisfy $\delta \in [1/2 - 1/2\sqrt{1-\exp(-4N\vert \alpha \vert^{2})} , 1/2 - 1/2\sqrt{1-\exp(-4\vert \alpha \vert^{2})}]$, with the infimum (supremum) of the $n_{\text{eff}}$ inequality enforced by the supremum (infimum) of the $\delta$-interval. It is important to address the $\delta$-dependence of this cat size measure, due the fact that there are examples of superpositions for which the cat size above has been shown to give ambiguous results depending on the value of $\delta$.\cite{dur}
 
Eq.(\ref{eqn:beam}) allows for a reinterpretation of the effective size in Eq.(\ref{eqn:helstrom}), which was originally defined for finite dimensional systems and required an integer effective size. In the present reformulation, $n_{\text{eff}}(\delta , \ket{\Omega})$ is defined for arbitrarily small $\delta$ and can be any real number. It is clear that given $\delta$, the entangled coherent state $\ket{\Omega}$ in Eq.(\ref{eqn:beam}) with $N=n_{ \text{eff}}(\delta , \ket{\Omega})$ (integer) modes can be obtained from mixing a single-mode superposition $\propto  \ket{\sqrt{n_{\text{eff}}(\delta , \ket{\Omega})}\alpha} + \ket{-\sqrt{n_{\text{eff}}(\delta , \ket{\Omega})}\alpha}$ with $n_{\text{eff}}(\delta, \ket{\Omega})-1$ vacuum modes. The maximal probability over all single mode measurements for distinguishing the branches of the latter single-mode superposition is $1-\delta$, the same as the probability for distinguishing the branches of the $N$-mode entangled coherent state with $N-n_{\text{eff}}(\delta , \ket{\Omega})$ modes traced over. Hence, the following revised definition for cat size allows one to consider both an arbitrary success parameter and a noninteger effective size:

\underline{Definition $\tilde{1}$}:  Given  $0<\delta <1/2$, the cat size $\tilde{C}_{\delta}(\ket{\Omega})$ of $\ket{\Omega} \propto \ket{\alpha}^{N}+\ket{-\alpha}^{N}$ is \begin{equation} \tilde{C}_{\delta}(\ket{\Omega}) = {N\over \tilde{n}_{\text{eff}}(\delta , \ket{\Omega})} \end{equation} where {\footnotesize \begin{equation} \tilde{n}_{\text{eff}}(\delta, \ket{\Omega}) = \text{min}_{n \in \mathbb{R}_{+}}\Bigg\lbrace {1\over 2} +{1\over 4}\Vert \; \ket{\sqrt{n}\alpha} \bra{\sqrt{n}\alpha}  - \ket{-\sqrt{n}\alpha} \bra{-\sqrt{n}\alpha} \; \Vert_{1}> 1-\delta \Bigg \rbrace  \end{equation}}

Notably, $\tilde{C}_{\delta}(\ket{\Omega})$ depends on both the number of modes $N$ involved in the entangled coherent state and also on the intensity $\vert \alpha \vert^{2}$ of the single-mode field in each branch (i.e. the expected photon number). We will see that an effective superposition size which depends solely on either the number of modes in the superposition or on the number of particles will fail to apply in general to superpositions with some macroscopic character.

An important case of the measurement-based cat size of Definitions $1$, $\tilde{1}$ occurs for a single-mode superposition of coherent states ($N=1$); this case is the most common setting for studies of photonic cat states to date. We see from Eq.(\ref{eqn:cat}) that $C_{\delta}(\ket{\Omega(N=1)}) \sim \vert \alpha \vert^{2}$, as expected from decoherence studies (see below). However, it should be noted that the same scaling can be obtained when one restricts the allowed measurements used to distinguish the branches to ``realistic" measurements, \textit{e.g.}, projective measurements in a classical or quasi-classical pointer state basis.\cite{gisin} Instead of the trace distance, the $L^{1}$ distance between the classical probability distributions $p(\xi \vert \rho_{A(B)})$ resulting from the chosen POVM $\lbrace E(\xi) \rbrace_{\xi \in \mathbb{R}}$ can then be used as the distinguishability metric.

In Definitions $1$ and $\tilde{1}$, we have neglected the presence of unbounded operators on $\ell^{2}(\mathbb{C})^{\otimes N}$. Defining the matrix elements of the $n-$RDM using $n$-element multi-indices $\text{\textbf{i}}$, $\text{\textbf{j}}$: \begin{equation}(\rho_{A/B}^{(n)})_{\text{\textbf{j}}}^{\text{\textbf{i}}}: = {1\over \text{tr}\rho_{A/B}^{(n)}}\text{tr}(a^{\dagger}_{i_{1}}\cdots a^{\dagger}_{i_{n}}a_{j_{1}}\cdots a_{j_{n}}\rho_{A/B} ) \end{equation} one could apply the effective size measure of Eq.(\ref{eqn:helstrom}), based on the condition that an $n$-photon measurement was made on the system instead of the condition of $n$ modes having been traced over. However, this definition is not useful for our purposes (\textit{i.e.} to determine an effective size for $\ket{\Omega}$), because it is clear that the $n$-RDMs of the two branches of $\ket{\Omega}$ have the same matrix elements for all $n$. Physically this is due to the fact that every $n$-photon measurement outcome occurs with the same probability in a product of coherent states as in a product of the same coherent states that have been $\pi$-rotated. It is also clear that any $n$-photon measurement leaves $\ket{\Omega}\bra{\Omega}$ invariant; there is no cat collapse from these measurements. We will return to the problem of using unbounded operators in optimal measurements for determining effective superposition size in our discussion of the relative quantum Fisher information below. This will be an essential component of showing that superpositions of the form $\ket{\Omega}$ are ``larger" than the largest cat-states of $(\mathbb{C}^{2})^{\otimes N}$ spin systems, $\ket{\text{GHZ}_{N}}$.

More evidence for $\mathcal{O}(N\vert \alpha \vert^{2})$ scaling of the cat size of $\ket{\Omega}$ based on branch distinguishability is provided by the two-state superposition size measure of Marquardt, Abel, and von Delft.\cite{marquardt} Let $\ket{\psi} \propto \ket{A} + \ket{B} \in \mathcal{K}^{\otimes N}$ (with $\mathcal{K}$ a single-mode Hilbert space) be a such a state. Define the 1-D Hilbert space $\mathcal{H}_{0} := \mathbb{C}\ket{A} \subset \mathcal{K}^{\otimes N}$ and define the Hilbert space $\mathcal{H}_{n}$ recursively by: \begin{eqnarray}\label{eqn:recursive}
\mathcal{H}_{n}&:=& \text{span }\lbrace \ket{v^{(n)}_{\xi , ij}} \big \vert \ket{\xi}\in \mathcal{H}_{n-1} \rbrace \nonumber \\ \ket{v^{(n)}_{\xi , ij}} &\propto & \left( a_{i}^{*}a_{j}  - \bigvee_{\ell = 0}^{n-1}P_{\mathcal{H}_{\ell}} \right) \ket{\xi}
\end{eqnarray}
where $P_{\mathcal{H}_{i}}$ is the projection onto the $i$-th Hilbert space and ``$\vee$" is the join of the lattice of these projections. Defining an orthonormal basis for $\mathcal{H}_{k}$ by $\lbrace \ket{ e^{(k)}_{i} } \rbrace $ (where $\ket{ e^{(0)} } = \ket{A}$), one can write $\ket{B} = \sum_{d,i} \lambda^{(d)}_{i} \ket{e^{(d)}_{i}} $ so that the probability $p(d)$ of obtaining the outcome of a projective measurement of $\ket{B}$ onto $\mathcal{H}_{d}$ is $\sum_{i}\vert \lambda^{(d)}_{i} \vert^{2}$. We define by $s$ the expected value of the Hilbert space label of the outcome of the countable POVM $\lbrace P_{\mathcal{H}_{\ell}} \rbrace_{\ell=0}^{\infty}$ (applied to $\ket{B}$). The Hilbert space containing the expected result of this projective measurement can be reached from $\ket{A}$ by application of $s$ single-photon operations of the form $a_{j}^{\dagger}a_{i}$. Although the $s$ calculated in the process of obtaining $\ket{B}$ from $\ket{A}$ can, in general, be different from the analogous quantity calculated in the opposite direction, \cite{marquardt, dur} we consider the $\ket{A}\rightarrow \ket{B}$ procedure carried out on $\ket{\Psi}$ (see next paragraph) to be instructive in demonstrating how this method works in infinite-dimensions.

Let $\ket{\Psi} \propto \ket{0}^{\otimes N} + \ket{2\alpha}^{\otimes N}$ and append an auxiliary mode (with states denoted by $\ket{\cdot}_{(0)}$) containing a coherent state of amplitude $\beta$ to the system so that the cat state becomes $\ket{\beta}_{(0)}\ket{0}^{\otimes N} + \ket{\beta}_{(0)}\ket{2\alpha}^{\otimes N}$. The auxiliary mode is necessary so that not all single-photon operators $a_{j}^{\dagger}a_{i}$ vanish on the vacuum branch, $\ket{0}^{\otimes N}$, of the superposition. Set the first Hilbert space $\mathcal{H}_{0} = \mathbb{C}\ket{\beta}_{(0)}\ket{0}^{\otimes N}$. Following the procedure outlined above, one finds for the $n$-th Hilbert space: \begin{equation}
\mathcal{H}_{n} = \text{span }\lbrace \ket{\beta}_{(0)}\ket{a_{1}}\cdots \ket{a_{N}} \big\vert a_{i}\in \mathbb{Z} \; \forall i \; , \; \sum_{i=1}^{N}a_{i} = n \rbrace
\end{equation}

Using the expansion of products of coherent states in the Fock basis of $\ell^{2}(\mathbb{C})^{\otimes N}$, a projective measurement of $\ket{\beta}_{(0)}\ket{2\alpha}^{\otimes N}$ with outcome contained in $\mathcal{H}_{d}$ is found to occur with probability: \begin{eqnarray}
p(d) &=& e^{-N\vert \alpha \vert^{2}}\sum_{a_{1}+\ldots +a_{N}=d}{\vert \alpha \vert^{2a_{1}}\cdots \vert \alpha \vert^{2a_{N}}\over a_{1}!\cdots a_{N}!}\nonumber \\ &=&  e^{-N\vert \alpha \vert^{2}}{{(N\vert \alpha \vert^{2}})^{d}\over d!}
\end{eqnarray}

which is a Poisson distribution with parameter $s=N\vert \alpha \vert^{2}$. To show that this $s$ value can be carried over from $\ket{\Psi}$ to $\ket{\Omega}$, we must remove the constraint of generating $\mathcal{H}_{n}$ from $\mathcal{H}_{n-1}$ by using photon number-conserving operations. $\ket{\beta}_{(0)}\otimes \ket{\Psi}$ is mapped to $\ket{\beta}_{(0)}\otimes \ket{\Omega}$ by the unitary operator $\mathcal{D}=\mathbb{I}_{(0)}\otimes \bigotimes_{i=1}^{N}D_{i}(-\alpha)$.  In particular, the unitarity of $\mathcal{D}$ implies that $\ket{\beta}_{(0)}\otimes \ket{2\alpha}^{\otimes N}$ has the same amplitudes on $\mathcal{H}_{n}$ as $\mathcal{D}(\ket{\beta}_{(0)}\otimes \ket{2\alpha}^{\otimes N}) = \ket{\beta}_{(0)}\otimes \ket{\alpha}^{\otimes N}$ has on $\mathcal{D}\mathcal{H}_{n}$. Explicitly, we now have a modification of Eq.(\ref{eqn:recursive}) to: \begin{widetext} \begin{eqnarray} \label{eqn:recursive2}
\mathcal{D}\mathcal{H}_{0} &=& \mathbb{C}\ket{\beta}_{(0)}\ket{-\alpha}^{\otimes N} \nonumber \\ \mathcal{D}\mathcal{H}_{1} &=& \text{span}_{\mathbb{C}}\big\lbrace \ket{\beta}_{(0)} \ket{-\alpha} \cdots D_{i}(-\alpha)a_{i}^{\dagger}\ket{0} \cdots \ket{-\alpha} \big\vert i=1,\ldots ,N \big\rbrace  \nonumber \\
{} &\vdots & {} \nonumber \\ \mathcal{D}\mathcal{H}_{n}&:=& \text{span }\lbrace \ket{\tilde{v}^{(n)}_{\xi , ij}} \big \vert \ket{\xi}\in \mathcal{D}\mathcal{H}_{n-1} \rbrace \nonumber \\ \ket{\tilde{v}^{(n)}_{\xi , ij}} &\propto & \left( \left( D_{j}(-\alpha)a_{j}D_{j}(\alpha) \right)^{\dagger}D_{i}(-\alpha)a_{i}D_{i}(\alpha)  - \bigvee_{\ell = 0}^{n-1}P_{\mathcal{D}\mathcal{H}_{\ell}} \right) \ket{\xi}
\end{eqnarray} \end{widetext}

In Eq.(\ref{eqn:recursive2}), the commutation relations $[ a^{\dagger} , D(\alpha)] = \overline{\alpha}D(\alpha)$, $[ a , D(\alpha)] = \alpha D(\alpha)$ of the displacement operators and creation/annihilation operators can be used to the effect of showing that we are now considering $(a_{j}^{\dagger}+\overline{\alpha}\mathbb{I}_{j})(a_{i}+\alpha\mathbb{I}_{i})$ to be the operators which when composed $s = N\vert \alpha \vert^{2}$ times takes the $\ket{\beta}_{(0)}\otimes \ket{-\alpha}^{\otimes N}$ branch of $\ket{\beta}_{(0)}\otimes \ket{\Omega}$  to the Hilbert space of the expected outcome of the POVM $\lbrace P_{\mathcal{D}\mathcal{H}_{i}} \rbrace$ on the $\ket{\beta}_{(0)}\otimes \ket{\alpha}^{\otimes N}$ branch.

It is clear that the recursive procedure defined in Eq.(\ref{eqn:recursive}) and Eq.(\ref{eqn:recursive2}) depends on both the number of modes, $N$, and the number of photons in the system. Due to the simple structure of the $\mathcal{H}_{i}$, we can interpret the parameter $s$ as the expected number of single photons required to transfer from the auxiliary coherent state and add to $\ket{0}^{\otimes N}$ to get a large overlap with $\ket{2\alpha}^{\otimes N}$. In Ref.[\onlinecite{marquardt}], $\ket{A}$ and $\ket{B}$ were assumed to have the same number of particles. Here we see that in passing to the infinite-dimensional Hilbert space associated with optical cavities, this measure can be useful in spite of having indefinite photon number in the cat state; in fact, we exploited an auxiliary coherent cavity as a source of an arbitrary number of single photons. The absence of a particle number superselection rule for massless bosons enables this procedure.

\subsubsection{\label{sec:fisher}Relative quantum Fisher information measure}
For an equal superposition of two states, the relative quantum Fisher information provides a measurement-based measure of effective cat size in spin systems which accounts for the number of modes involved in the superposition.\cite{dur} The generic system of interest in Ref.[\onlinecite{dur}] consisted of a finite number of spin-1/2 particles, so there was no difference between considering particle number and mode number. This effective superposition size is defined for a superposition $\ket{\psi}$ of orthogonal states $\ket{\psi_{0}}$, $\ket{\psi_{1}} \in (\mathbb{C}^{2})^{\otimes N}$ by:

\underline{Definition 2}: The relative quantum Fisher information effective size, $N_{\text{eff}}^{\text{rF}}$, for $\ket{\psi}={1/ \sqrt{2}} \left( \ket{\psi_{0}}+\ket{\psi_{1}} \right) \in (\mathbb{C}^{2})^{\otimes N}$ is: \begin{equation}\label{eqn:relativefish} N_{\text{eff}}^{\text{rF}}(\ket{\psi}) := { N_{\text{eff}}^{\text{F}}(\ket{\psi}) \over {1\over 2} N_{\text{eff}}^{\text{F}}(\ket{\psi_{0}}) + {1\over 2} N_{\text{eff}}^{\text{F}}(\ket{\psi_{1}})} \end{equation}

where $N_{\text{eff}}^{\text{F}}(\rho)= {1\over 4N}\text{max}_{\mathcal{A}} \mathcal{F}(\rho, A)$, $\mathcal{A} =  \lbrace \sum_{i=1}^{N}A^{(i)}\otimes 1^{\otimes N\setminus \lbrace i \rbrace }\, , A^{(i)}>0 \, , \Vert A^{(i)} \Vert =1 \rbrace$ (bounded, positive, local operators), and $\mathcal{F}(\rho, A)$ is the quantum Fisher information \cite{caves} for an \textit{arbitrary} state $\rho$ of $(\mathbb{C}^{2})^{\otimes N}$ evolved in time by $A$.

We will consider all $A$ to be time-independent, although this is not the most general situation. It is useful to note that for a pure state $\rho$, the quantum Fisher information for $\rho(t) = \exp (-iAt) \rho \exp (iAt)$ is: \begin{equation} \mathcal{F}(\rho(t)) =\mathcal{F}(\rho ,A) = 4\text{tr}(\rho (\Delta A)^{2}) \end{equation} which is proportional to the variance of the operator $A$ in the initial state $\rho$.  A cat state $\ket{\psi}$ is considered macroscopic if $N_{\text{eff}}^{\text{rF}}(\ket{\psi}) \in \mathcal{O}(N)$. Note that $N_{\text{eff}}^{\text{F}}$ alone can be used as a measure of macroscopicity for general quantum states which are not necessarily Schr\"{o}dinger cat states.\cite{dur}

This measure can be fully generalized to states of $\ell^{2}(\mathbb{C})^{\otimes N}$ by introducing unbounded operators on the Hilbert space into the maximization of quantum Fisher information in the numerator and denominator of Eq.(\ref{eqn:relativefish}). Restricting the maximization to the bounded operators $\mathcal{B}(\ell^{2}(\mathbb{C})^{\otimes N})$ ($\mathcal{B}(\mathcal{H})$ signifying bounded operators on the Hilbert space $\mathcal{H}$) allows lower bounds for $N_{\text{eff}}^{\text{rF}}$ to be derived. Thus the entangled coherent state in Eq.(\ref{eqn:entangledstate}) can then be shown to be macroscopic according to the relative quantum Fisher information measure by using these lower bounds. In particular, all that is required to derive a lower bound implying macroscopicity is to find a local operator $A\in \mathcal{A}$ which gives a variance of $\mathcal{O}(N^{2})$ in $\ket{\Omega}$. This follows from the fact that for a product state (\textit{e.g.}, each branch of $\ket{\Omega}$), optimizing over local, positive, bounded operators results in a maximal quantum Fisher information in $\mathcal{O}(N)$, \cite{smerzi} so that the denominator of Eq.(\ref{eqn:relativefish}) is $\mathcal{O}(1)$. An example of an operator giving the required scaling of variance in $\ket{\Omega}$ is $A=\sum_{i}A^{(i)}$ with $A^{(i)} = {1\over \sqrt{1-\exp{-4\vert \alpha \vert^{2}}}}\left( \ket{\alpha}\bra{\alpha} - \ket{-\alpha}\bra{-\alpha} \right)_{i}$, which for large $\vert \alpha \vert $, \textit{i.e.} neglecting the overlap of diametrically opposed coherent states, is analogous to $\sigma_{z}$ for a two-level systems while $\ket{\Omega}$ in this situation is analogous to an eigenvector of $\sigma_{x}^{\otimes N}$ with eigenvalue 1. One can verify directly that $A^{(i)}$ has operator norm 1, so $A$ has operator norm $N$.  Calculating the variance of $A=\sum_{i}A^{(i)}$ in $\ket{\Omega}$ results in the following inequality: \begin{equation} N_{\text{eff}}^{\text{rF}}(\ket{\Omega}) \ge N\left({1-e^{-4\vert \alpha \vert^{2}} \over 1+e^{-2N\vert \alpha \vert^{2}}}\right) + {e^{-2N\vert \alpha \vert^{2}}+ e^{-4\vert \alpha \vert^{2}} \over 1 +e^{-2N \vert \alpha \vert^{2}} } \end{equation} the right hand side of which is in $\mathcal{O}(N)$. Hence $\ket{\Omega}$ can be considered macroscopic by this measure. Note that the exact value of $N_{\text{eff}}^{\text{rF}}(\ket{\Omega})$ could be larger since we have not carried out the optimization over $A\in \mathcal{A}$ explicitly.

In the above calculation, we have neglected the creation and annihilation operators $a^{\dagger}_{i}$, $a_{i}$ on $\ell^{2}(\mathbb{C})^{\otimes N}$ (which are unbounded) and hence also the product of Heisenberg algebras spanned by $\lbrace a^{\dagger}_{i},a_{i},\mathbb{I}_{i} \rbrace$. Of particular interest in this algebra are the self-adjoint quadrature operators $x^{(\phi)}_{j}:=(a_{j}\exp -i\phi + a^{\dagger}_{j} \exp i\phi )/ \sqrt{2}$. Our neglect of these unbounded, local, self-adjoint operators in the set $\mathcal{A}$ of time-evolution generators used to calculate the relative quantum Fisher information effective size is apparently at fault for the lack of any polynomial scaling of the relative quantum Fisher information superposition size of $\ket{\Omega}$ with the photon number. It is clear from Eq.(\ref{eqn:relativefish}) in Definition 2 that changing the bounds of the local operators in $\mathcal{A}$ to some finite value greater than 1 does not change the relative quantum Fisher information effective size. However, if the (unbounded) quadrature operators are included in the optimization, the effective size can be larger. For instance, the variance of $\sum_{i}x^{(\phi)}_{i}$ in $\ket{\pm \alpha}^{\otimes N}$ is $N/2$, while the variance of the same operator in $\ket{\Omega}$ is at most $N^{2}\vert \alpha \vert^{2}\tanh N\vert \alpha \vert^{2} + N\vert \alpha \vert^{2} + {1\over 2}$. Hence, when the local operators defining elements of $\mathcal{A}$ are taken in $\mathcal{B}(\ell^{2}(\mathbb{C})^{\otimes N}) \cup \lbrace x^{(\phi)} \rbrace $, a larger lower bound to the effective size scaling for the entangled coherent state $\ket{\Omega}$ is achieved: \begin{equation} N_{\text{eff}}^{\text{rF}}(\ket{\Omega}) \ge N\vert \alpha \vert^{2}\tanh N\vert \alpha \vert^{2} + \vert \alpha \vert^{2} + {1\over 2N} \in \mathcal{O}(N\vert \alpha \vert^{2}) \end{equation} The $N\vert \alpha \vert^{2}$ scaling is important to consider because $\log_{N}(\vert \alpha \vert^{2}) $ is potentially much greater than 1. In addition, one could have chosen to append the local photon number operators $\sum_{i}a^{\dagger}_{i}a_{i}$ to the expanded set $\mathcal{A}$ above; including it results in the same $\mathcal{O}(N\vert \alpha \vert^{2})$ scaling of effective cat size. We have not optimized the variances in the numerator and denominator over every possible sum of local operators acting on $\ell^{2}(\mathbb{C})^{\otimes N}$; \textit{i.e.}, we have not considered all $A=\sum_{i}u^{(i)}$ where $u^{(i)}$ is in the universal enveloping algebra of the $i$-th mode. According to the analysis above and the definition of quantum Fisher information, we can say only that there are certain local evolutions for this system for which the square of the Bures velocity\cite{caves} $(ds_{\text{Bures}}/dt)^{2}$ along the local evolution is greater in the superposition $\ket{\Omega}$ than its branches by a factor of $\vert \alpha \vert^{2}$. 

A similar analysis of relative quantum Fisher information can be used to examine the effective size scaling of a ``hierarchical cat state," which is an $\ell^{2}(\mathbb{C})^{\otimes N}$ analog of the $\ket{\text{GHZ}_{N}}$ spin state. However, this state exhibits richer internal structure than $\ket{\text{GHZ}_{N}}$, as each of its branches is composed of smaller Schr\"{o}dinger ``kittens": \begin{equation}\label{eqn:hierarchical}  \ket{\text{HCS}_{N}(\alpha)} := {1\over \sqrt{2}}\left(\left( {\ket{\alpha}+\ket{-\alpha}\over A_{+}(\vert \alpha \vert)} \right)^{\otimes N}+ \left( {\ket{\alpha}-\ket{-\alpha}\over A_{-}(\vert \alpha \vert) }\right)^{\otimes N}\right) \end{equation}

Restricting the maximization over local, self-adjoint operators to bounded operators yields an effective size of $N$ for $\ket{\text{HCS}_{N}(\alpha)}$ for all $\alpha \neq 0$, just as for any $\ket{\text{GHZ}_{N}}$ state. This value is also predicted by all measurement-based measures considered in this paper and also by the comparison-based measures of D\"{u}r, Simon, and Cirac to be discussed in Section \ref{sec:comparison}. Physically, $\ket{\text{HCS}(\alpha)}$ is a superposition of parity eigenstates, with one branch a product state of ``even kittens" or a product state of ``odd kittens". When considered in light of the $\mathcal{O}(N\vert \alpha \vert^{2})$ scaling of $N^{\text{rF}}_{\text{eff}}(\ket{\Omega})$, one expects from the analyses of Ref.[\onlinecite{cirac}] of that the hierarchical cat state should exhibit an effective superposition size at least as large as $\ket{\Omega}$ because its branches are orthogonal and it is in the same Hilbert space as $\ket{\Omega}$. However, proceeding with our earlier approach of introducing the local quadrature operators or even the local photon number operators into the maximization of quantum Fisher information gives only a smaller lower bound: $N^{\text{rF}}_{\text{eff}}(\ket{\text{HCS}(\alpha)}) \ge 1$. From additionally considering local operators in $\mathcal{A}$ of the form $A=\sum_{i}a^{\dagger}_{i}\sigma^{(i)}a_{i}$, where $\sigma^{(i)}$ can be $\sigma^{z}_{i}$ or $\sigma^{x}_{i}$, it is found that $N^{\text{rF}}_{\text{eff}}(\ket{\text{HCS}(\alpha)})$ is at least $\mathcal{O}(N)$, but no dependence on $\vert \alpha \vert$ is observed. This result is due to fact that the branches have quantum Fisher information scaling as the same power of $\vert \alpha \vert$ as the superpositon itself when these operators are included in the maximization, so the powers cancel in Eq.(\ref{eqn:relativefish}). Although we do not yet have a precise estimate of the relative quantum Fisher information effective size for $\ket{\text{HCS}_{N}(\alpha)}$, we find no reason to believe that it should scale as $\mathcal{O}(N\vert \alpha \vert^{k})$ with $k > 0$. The consequence of $\mathcal{O}(N)$ scaling of $\ket{\text{HCS}_{N}}$ is that in an infinite-dimensional Hilbert space, there are superpositions (\textit{e.g.}, $\ket{\Omega}$) which have larger superposition size than the continuous variable analogs of $\ket{\text{GHZ}_{N}}$ states. Because $\ket{\Omega}$ is an analog of $\ket{\text{GHZ}_{N}(\epsilon)}$ in spin systems, we obtain the principal result of this paper: in $\ell^{2}(\mathbb{C})$ there are superpositions with nonorthogonal branches which have larger effective sizes than superpositions with orthogonal branches.

\begin{figure}
\includegraphics[scale=.5]{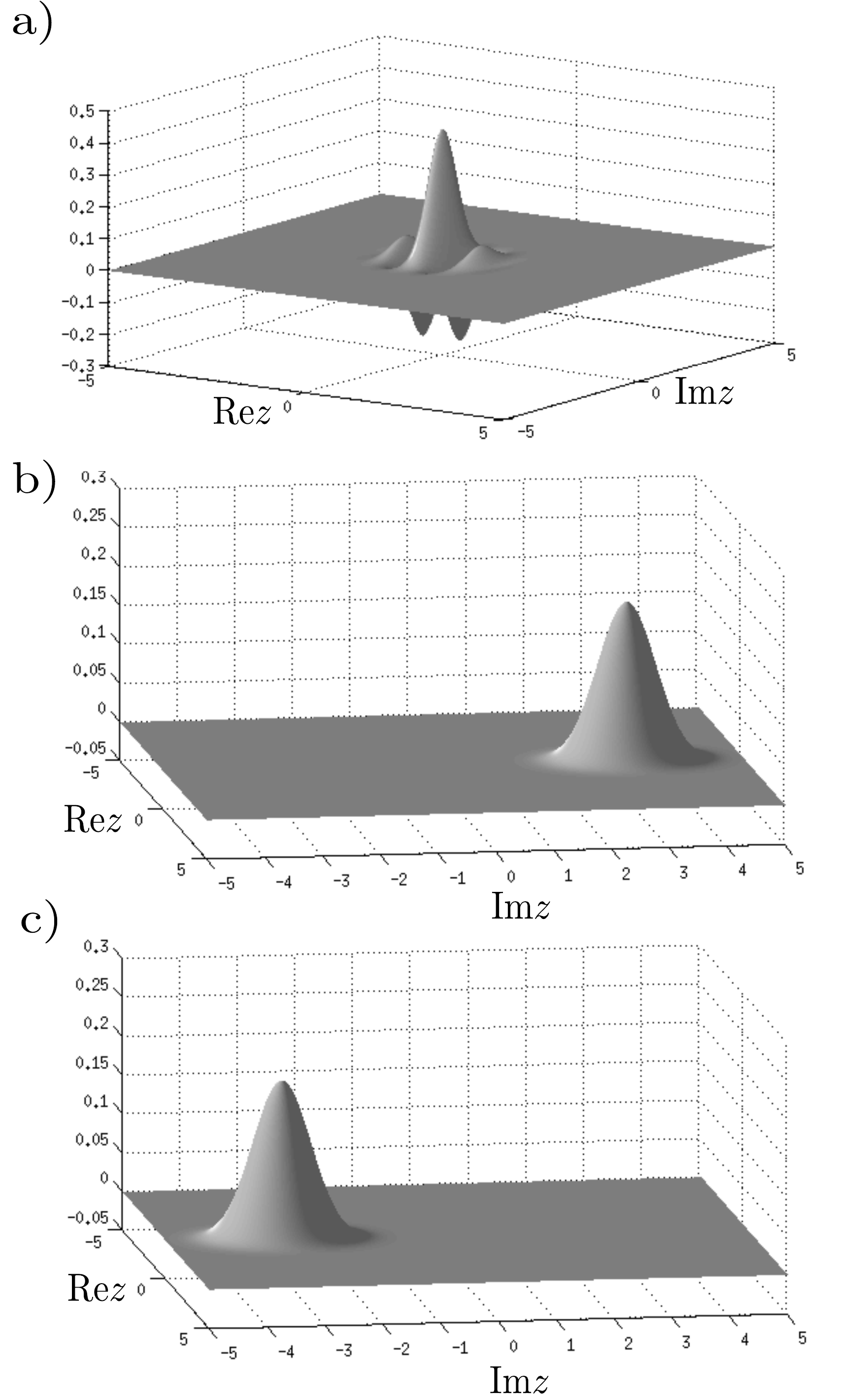}
\vspace{-0.2cm}
\caption{\label{fig:wigner} Graphs of the Wigner function $W^{\text{HCS}_{2}(\alpha)}(\gamma_{1},\gamma_{2})$ ($(\gamma_{1} , \gamma_{2}) \in \mathbb{C} \times \mathbb{C}$) of $\ket{\text{HCS}_{2}(\alpha)}$ (with $\vert \alpha \vert = 3$) for a) $\gamma_{2}=0$, b) $\gamma_{2}=3$, c) $\gamma_{2}=-3$. See Eq.(\ref{eqn:wignerhcs}) in Appendix \ref{sec:wignerhcs2} for the explicit form of this function on $\mathbb{C}\times \mathbb{C}$. }
\end{figure}

The lack of maximally ``cat-like" characteristics of $\ket{\text{HCS}_{N}(\alpha)}$ is also demonstrated by the Wigner function $W:\mathbb{C}\times \mathbb{C} \rightarrow \mathbb{R}$ for the state $\ket{\text{HCS}_{2}(\alpha)}$, which exhibits a peak at the origin of $\mathbb{C}\times \mathbb{C}$ in addition to smaller peaks at $(\pm \alpha , \pm \alpha)$ (see Figure \ref{fig:wigner} and Appendix \ref{sec:wignerhcs2} for the explicit function). Calculation of the general $N$-mode Wigner function shows that the peak at the origin persists for $N>2$. In contrast, the $N$-mode Wigner function of $\ket{\Omega}$ exhibits only two peaks on $\mathbb{C}^{N}$ separated by $2\sqrt{N}\vert \alpha \vert$ (see Section \ref{sec:intrinsic}). It should also be noted that $\ket{\text{HCS}_{N}(\alpha)}$ can be written (in the same basis) as a superposition with many branches, so it is not a two-branch superposition in the original spirit of Schr\"{o}dinger's cat.

\subsection{\label{sec:comparison}Comparison-based measures of cat size}

In contrast to a measurement-based definition of superposition size, a comparison-based definition requires a reference state which has a known size under that definition. For  entangled qubit systems, all superposition size measures discussed in this work concur on a maximal superposition size of $N$, which occurs for the $\ket{\text{GHZ}_{N}}$ state. We can extend previous comparison-based superposition size measures developed for spin-1/2 systems \cite{cirac} to the state in Eq.(\ref{eqn:entangledstate}) by restricting to a 2D subspace of the single-mode Hilbert space. To apply the distillation protocol implemented in Ref.[\onlinecite{cirac}], one identifies an operator $E_{1}= k[ \vert e_{1} \rangle \langle e_{2} \vert + \vert e_{2} \rangle \langle \varphi_{-} \vert ]$ (acting on the same $\mathbb{C}^{2}$ subspace of $\ell^{2}(\mathbb{C})$ as introduced in Section \ref{sec:korsbakken})in which $\langle \varphi_{-}\vert -\alpha \rangle =0$ and $k$ is chosen such that the complementary element $E_{2}^{\dagger}E_{2}$ in the POVM $\lbrace E_{1}^{\dagger}E_{1}, E_{2}^{\dagger}E_{2} \rbrace $ is rank one. With $k$ chosen appropriately, $E_{2}$ will have the form ${1\over \Vert \ket{\chi} \Vert} \vert \chi \rangle \langle \chi \vert $. Note that if $E_{1}$ is applied to each mode of $\ket{\Omega}$, a GHZ state $\propto \ket{e_{1}}^{\otimes N} + \ket{e_{2}}^{\otimes N}$ will result whereas $E_{2}$ applied to each mode results in a product state. The goal is to find the expected value of $n(N)$ when applying POVM above to each mode, transforming $\ket{\Omega}$ to $(1/ \sqrt{2})\ket{\chi}^{\otimes N-n(N)}\otimes \left( \ket{e_{1}}^{\otimes n(N)} + \ket{e_{2}}^{\otimes n(N)} \right)$. The only subtlety in this computation is that the first successful $E_{1}$ outcome changes the normalization of the post-measurement state; after the first $E_{1}$ outcome, the measurements are binomially distributed: the probability of an $E_{1}$ outcome being $1-\exp(-2\vert \alpha \vert^{2})$ ($\exp(-2\vert \alpha \vert^{2})$ for an $E_{2}$ outcome). The probability of the first $E_{1}$ outcome occurring at the $m$-th measurement is \begin{equation}
p_{m} = e^{(N-2m+1)\vert \alpha \vert^{2}} {\sinh(\vert \alpha \vert^{2}) \over \cosh(N \vert \alpha \vert^{2})}
\end{equation}

The probability of the final state having $n(N) = n$ is \begin{equation} \binom{N}{n}e^{-(N-1)\vert \alpha \vert^{2}} (e^{2\vert \alpha \vert^{2}}-1) {\sinh(\vert \alpha \vert^{2}) \over \cosh(N \vert \alpha \vert^{2})} \end{equation} so $\langle n(N) \rangle = N{1-e^{-2\vert \alpha \vert^{2}} \over 1+e^{-2N\vert \alpha \vert^{2}}}$. The distillation protocol applied to entangled coherent states is not very useful; an entangled coherent state has an integer number of modes and the above result indicates the unsurprising fact that for $\vert \alpha \vert $ even moderately large, the $N$-mode $\ket{\text{GHZ}_{N}}$ state \begin{equation} {1\over \sqrt{2}}\left( \ket{\alpha}^{\otimes N} + \ket{e_{2}}^{\otimes N} \right) \label{eqn:newghz}\end{equation} can be distilled with very low error from $\ket{\Omega}$ due to the near orthogonality of the branches. According to this comparison-based size, the state in Eq.(\ref{eqn:ghz}) is equivalent to the same state except with the single particle states replaced by single-mode states $\ket{\alpha}$ and $\ket{e_{2}}$-- the structure of the modes has not been considered. In addition, it is not clear how the distillation process affects the cat size. What is perhaps interesting about this procedure is that the state in Eq.(\ref{eqn:newghz}) can be transformed into the ``hierarchical" cat state of Eq.(\ref{eqn:hierarchical}) by local unitaries. Hence Eq.({\ref{eqn:beam}}) provides a simple method for using beamsplitters, phase shifters, and local measurements to generate these states from mixing of a single-mode coherent state superposition with vacuum.

Another comparison-based measure of cat size for the entangled coherent state comes from comparing the magnitude of off-diagonal elements of the $n$-RDM of standard GHZ$_{N}$ states to those of the $n$-RDM of entangled coherent states after probabilistic loss of $N-n$ modes. \cite{cirac} Consider the decision between tracing over a mode of Eq.(\ref{eqn:ghz}) with (small) probability $\lambda$ and leaving it alone with probability $1-\lambda$. If this decision is made for each of the $N$ modes, the expected value of the off-diagonal elements of the density matrix is ${1\over 2}(1-\lambda)^{N} \approx {1\over 2}e^{-N\lambda}$ because if a decision is made to trace over any mode, the resulting state is mixed. In same procedure for $\ket{\Omega}$ one expects to trace over $N\lambda$ modes. Taking into account the normalization factor of $\ket{\Omega}$, the expected off-diagonal amplitude is: \begin{equation} {e^{-2N\lambda \vert \alpha \vert^{2}}\over 2+2e^{-2N \vert \alpha \vert^{2}}} = {1\over 2} e^{-2N\lambda \vert \alpha \vert^{2} - \log (1+\exp(-2N\lambda \vert \alpha \vert^{2})) } \end{equation}

Up to a negligible logarithmic term, the expected off-diagonal amplitude of the entangled coherent state after probabilistic mode loss is the same as that of a GHZ$_{M}$ state with $M=2N\vert \alpha \vert^{2}$, under the assumption that $\vert \alpha \vert^{2} \in \mathbb{Z}_{+}$. This result for probabilistic mode loss can be compared to the change in the amplitude of the fringes of the $N$-mode Wigner function of $\ket{\Omega}$. Integrating over the phase space of $n<N$ modes results in a new Wigner function on $\mathbb{C}^{N-n}$ with the oscillating (i.e., non-Gaussian) term suppressed by $\exp (-n\vert \alpha \vert^{2}/2)$. One sees that the partial trace over a mode corresponds to tracing out all of the constituent particles and excitations in the mode.

The difference between considering particles versus modes in defining a cat size is precisely analogous to the difference between definitions of 1-RDMs in, \textit{e.g.} the study of Bose-Einstein condensation \cite{ueda} versus the definition employed in disucssions of qubit manipulations. \cite{nielsen} In the former, one takes $(\rho^{(1)}_{\Omega})_{ij} = \bra{\Omega} a_{i}^{\dagger}a_{j} \ket{\Omega}$, with $N^{2}$ matrix elements defined by modes. In the latter, one defines $(\rho^{(1)}_{\Omega})_{ij} = \left( \text{tr}_{2,\ldots ,N} \ket{\Omega}\bra{\Omega} \right)_{ij}$ with the matrix elements taken with respect to Fock states. The trace of the 1-RDM defined through particle creation/annihilation for $\ket{\Omega}$ is $N\vert \alpha \vert^{2} \tanh (N\vert \alpha \vert^{2})$; this value is equal to the expected value of the sum of number operators for each mode, while the trace of the mode-defined 1-RDM is unity. More drastic consequences of this distinction are encountered in the formulations of entanglement measures for indistinguishable particles, \cite{wiseman} for which the total Hilbert space is the symmetrized or antisymmetrized tensor product of mode Hilbert spaces. For the branch distinguishability measures considered in this paper, the importance of this distinction is restricted to bosonic systems; for fermions, the Pauli exclusion principle results in the equivalence of effective superposition sizes based on mode number and particle number for an appropriate definition of the modes. This fact has been used to derive an upper bound for the effective cat size (according to Definition 1) of superpositions of clockwise/anti-clockwise circulation states of superconducting flux qubits in terms of the sum of electron number differences for each mode comprising the circulation states: $\sum_{i} \langle \circlearrowleft \vert c^{\dagger}_{i}c_{i} \vert \circlearrowleft \rangle - \langle \circlearrowright \vert c^{\dagger}_{i}c_{i} \vert \circlearrowright \rangle$. \cite{korsbakken2009}

\subsection{\label{sec:intrinsic}Empirical measures of cat size}
For a single cavity, empirical definitions (i.e., requiring neither optimization over measurements nor comparisons to other states) of effective superposition size can be obtained from the Wigner function of the state itself. \cite{jeong} For example, the phase-space interference fringes of the Wigner function of the even superposition $\propto \ket{\alpha} + \ket{-\alpha}$, $\alpha \in \mathbb{R}_{+}$ are parallel to the imaginary axis and have wavelength $\pi / 2\alpha$. The distance between Wigner function peaks is $2\alpha$.  The effective size of Lee and Jeong \cite{jeong} takes both of these factors into account and was written with general multiple-mode quantum states in mind. Applied to the many-mode Wigner function for the state in Eq.(\ref{eqn:entangledstate}), the effective size consists of $N$ identical sums. The Wigner function $N$-mode entangled coherent state (an entire function from $\mathbb{C}^{N} \rightarrow \mathbb{R}$) has peaks separated by $2\sqrt{N}\alpha$. The square of this distance agrees with the scaling of our result in Eq.(\ref{eqn:cat}). Other intrinsic notions of cat size in terms of the decoherence time ${1/ \langle \hat{n} \rangle }$ \cite{harochedavidovich, harochetransact} of the cavity state have been introduced based on experimental results and simple models of nonunitary cavity state evolution. These notions are less well-defined for nonlocal superpositions, i.e. superpositions in the Hilbert space of many cavity modes. For instance, a description of decoherence of the hierarchical cat state Eq.(\ref{eqn:hierarchical}) must include the intermode decoherence time (the decoherence time of the device coupling the modes) and all of the intramode (cavity) decoherence times.  

\section{\label{sec:conclusion}Conclusion}

We have extended effective size measures of quantum superpositions of two classically distinguishable states (i.e. Schr\"{o}dinger cat states) from the spin state setting of $(\mathbb{C}^{2})^{\otimes N}$ to the most widely studied cat states of $\ell^{2}(\mathbb{C})^{\otimes N}$, which are the entangled coherent states (Eq.(\ref{eqn:entangledstate})). The measures fall into three classes: 1) those relating to a distinguishability problem requiring some optimization over positive operators or selection of an appropriate effect operators (the measurement-based measures), 2) those requiring comparison to a superposition state with fixed size (comparison-based measures), and 3) those defined from the time-evolution of geometric properties of the many-mode quasiprobability distribution  (empirical or intrinsic measures). The latter class of measures is most amenable to current experimental techniques because the optimal measurements defining the effective size in the measurement-based class of measures can be impractical or even impossible to realistically implement. However, the measurement-based measures give more insight into the quantum information aspects of large superpositions because they are expressed in terms of metrics on the space of density operators.

By extending these measures, we have shown that the cat size of entangled coherent states of the form in Eq.(\ref{eqn:entangledstate}) depends on both the number of modes ($N$) and intensity of the field in each mode (scaling as $\vert \alpha \vert^{2}$) as $N\vert \alpha \vert^{2}$. We have demonstrated that this scaling can be obtained only through effective size measures which take into account both the number of modes involved in the superposition (for instance, the number of resonant microwave cavities, each under the assumption of monochromaticity) and the number of particles (or intensity of photons) in each mode. These criteria are not fulfilled in, \textit{e.g.}, the distillation protocol of Ref.[\onlinecite{cirac}], and the relative quantum Fisher information measure of Ref.[\onlinecite{dur}] when unbounded operators are neglected in the maximization of quantum Fisher information.

Using the relative quantum Fisher information measure of effective size, we have shown that a hierarchical cat state of the form $\ket{\text{HCS}_{N}(\alpha)}:= 1/\sqrt{2} (\ket{\psi_{+}}^{\otimes N}+\ket{\psi_{-}}^{\otimes N})$ with $\ket{\psi_{\pm}} \propto \ket{\alpha} \pm \ket{-\alpha}$ fails to exhibit the enhanced scaling with mode intensity obtained for $\ket{\Omega}$, despite having orthogonal branches. This result was obtained by carrying out a maximization of the quantum Fisher information of the hierarchical cat state and its branches over the union of local, positive operators contained in $\mathcal{B}(\ell^{2}(\mathbb{C})^{\otimes N}) \cup \lbrace \sum_{i}x^{(\phi)}_{i} , \sum_{i}a_{i}^{\dagger}a_{i} \rbrace $ (bounded operators and an important set of unbounded operators, namely, the local quadrature and number operators). This suggests that $\ket{\text{HCS}_{N}(\alpha)}$, which is a $\ell^{2}(\mathbb{C})^{\otimes N}$ analog of a $\ket{\text{GHZ}_{N}}$ state but exhibits richer internal mode structure, is not the largest cat state in systems of entangled electromagnetic cavities. Because photons are the elementary excitations of the system under consideration, and the notion of superposition size based on single-photon decoherence in quantum phase space suggest decoherence rates for a superposition which are inversely proportional to the photon number, \cite{harochebook} we conjecture that the largest possible superposition size in $\ell^{2}(\mathbb{C})^{\otimes N}$ exhibits $\mathcal{O}(N\langle \sum_{i=1}^{N}a^{\dagger}_{i}a_{i} \rangle)$ scaling. Therefore, the entangled coherent states of the form of Eq.(\ref{eqn:entangledstate}) are among the largest possible superpositions in this space.

The study of superposition sizes and measures of quantum state macroscopicity is also relevant to fundamental physics. The superpositions of standard coherent states (both single-mode and many-mode) treated here are the simplest examples of non-local superpositions for continuous quantum variables. The development and application of cat size measures for superpositions of states occurring in exotic phases of matter (\textit{e.g.}, those allowing for superpositions of soliton/domain-wall solutions of the quantum equations of motion,\cite{zurek} or those containing superpositions of gauge field configurations\cite{polyakov}) could be useful for a general understanding of the conditions under which the principles of quantum mechanics may be extended to macroscopic systems.

\appendix
\section{\label{sec:app}Proof of Eq.(\ref{eqn:beam})}

Consider $M$ photon cavities with respective creation/annihilation operators $a^{\dagger}_{i},a_{i}$ and let $\mathcal{B}_{ij}(\theta) := P_{i}(\pi /2) B_{ij}(\theta)P_{j}(\pi /2)$ where $P_{i}(\phi) = e^{i\phi a^{\dagger}_{j}a_{j}}$ is a phase-shifter on the $i$-th mode and and $B_{ij}(\theta) = e^{i\theta a_{i}^{\dagger}a_{j} +a_{j}^{\dagger}a_{i} }$ is a beamsplitter acting on modes $i$ and $j$. Then we have the following:

\underline{Lemma} \begin{equation}\label{eqn:lemma} \ket{\alpha}^{\otimes M} = \prod_{q=1}^{M-1}\mathcal{B}_{q,q+1}(\theta_{q})\left(\ket{\sqrt{M}\alpha}_{1}\otimes \ket{0}^{\otimes M-1} \right) \end{equation} where $\theta_{q} = (\tan^{-1}\circ \sec)^{M-1-q}(\pi /4)$

\underline{Pf} Using the fact that for any two coherent states $\ket{\alpha}_{i}$, $\ket{\beta}_{j}$, $\mathcal{B}_{ij}(\theta)\ket{\alpha}_{i}\ket{\beta}_{j} = \ket{\alpha \cos \theta +\beta \sin \theta}_{i}\ket{\alpha \sin \theta - \beta \cos \theta}$, one can evaluate the right hand side of Eq.(\ref{eqn:lemma}) as : \begin{eqnarray}
{}&{}&\ket{\sqrt{M}\alpha \cos \theta_{1}}_{1} \otimes \ket{\sqrt{M}\alpha \sin\theta_{1} \cos \theta_{2}}_{2} \nonumber \\ &\otimes & \ldots \otimes \big\vert \sqrt{M}\alpha \prod_{i=1}^{M-2}(\sin \theta_{i}) \cos\theta_{M-1}\big\rangle_{M-1} \nonumber \\ &\otimes & \big\vert \sqrt{M}\alpha \prod_{i=1}^{M-2}(\sin \theta_{i}) \sin\theta_{M-1}\big\rangle_{M}
\end{eqnarray}

Each of the kets in this product is seen to be equal to $\ket{\alpha}$ by using the rules for composition of trigonometric functions and inverse trigonometric functions, e.g. in the $M-1$ ket:  the identity $\sin((\tan^{-1}\circ \sec)^{n}(\pi /4)) = \sin\left( \tan^{-1}\left( (\sec \circ \tan^{-1})^{n-1}(\sqrt{2})\right) \right)  = \sqrt{n+1 \over n+2}$ can be used to simplify to the product to $\alpha \sqrt{M} \cos\theta_{M-1} \prod_{i=1}^{M-2}\sqrt{M-i \over M-i+1} = \alpha {\sqrt{M}\over \sqrt{2}}\left({\sqrt{M-1}\over \sqrt{M}}  {\sqrt{M-2}\over \sqrt{M-1}} \ldots  {\sqrt{2}\over \sqrt{3}} \right)=\alpha$ These manipulations prove the lemma. $\square$

Eq.(\ref{eqn:beam}) follows from replacing $\ket{\sqrt{M}\alpha}_{1}$ in the first mode with the unnormalized superposition $\ket{\sqrt{M}\alpha}_{1}+\ket{-\sqrt{M}\alpha}_{1}$, following the proof of the Lemma with $\alpha \rightarrow -\alpha$, and normalizing the resulting state.

\section{\label{sec:appbranch}Optimal measurement for branch distinguishability}

We provide an explicit proof that the measurement which optimally distinguishes the branches of $\ket{\psi_{+}}$ considered as pure states ($\rho_{\alpha}=\ket{\alpha}\bra{\alpha}$, $\rho_{-\alpha} = \ket{-\alpha}\bra{-\alpha}$ present with equal \textit{a priori} probability) collapses $\ket{\psi}$ to $\ket{\pm \alpha}$ with high probability. We work in a basis $\lbrace \ket{\psi_{+}},\ket{\psi_{-}} \rbrace$ for a 2-D subspace of $\ell^{2}(\mathbb{C})$. The POVM which gives the minimal probability of error (maximal probability of success) for distinguishing $\rho_{\alpha}$ from $\rho_{-\alpha}$ consists of the projectors onto the 1-D eigenspaces of $\rho_{\alpha} - \rho_{-\alpha} = \sqrt{1-\exp(-4\vert \alpha \vert^{2})}\sigma^{x}$ and in this case is given by $\lbrace E_{i} = \ket{\xi_{i}}\bra{\xi_{i}} \rbrace_{i = \pm}$ with $\ket{\xi_{\pm}} = 1/ \sqrt{2} ( \ket{\psi_{+}} \pm \ket{\psi_{-}})$. The measurement results $E_{\pm}$ obtained on the pure cat state $\ket{\psi_{+}}\bra{\psi_{+}}$ occur with equal probability and because the optimal POVM consists of rank-1 projectors onto $\ket{\xi_{\pm}}$, the resulting state is (respectively) $\ket{\xi_{\pm}} \propto E_{\pm}^{1/2}\ket{\psi_{+}}\bra{\psi_{+}}E_{\pm}^{* \, 1/2}$ with probability $1/2$. Explicitly: \begin{eqnarray}
\ket{\xi_{\pm}} &=& \left( {1\over{2\sqrt{1+e^{-2\vert \alpha \vert^{2}}}}} \pm {1\over{2\sqrt{1-e^{-2\vert \alpha \vert^{2}}}}} \right) \ket{\alpha} \nonumber \\ &{+}&  \left( {1\over{2\sqrt{1+e^{-2\vert \alpha \vert^{2}}}}} \mp {1\over{2\sqrt{1-e^{-2\vert \alpha \vert^{2}}}}} \right) \ket{ - \alpha} 
\end{eqnarray}

It is clear that for any physical number of photons ($\vert \alpha \vert^{2} > 1$), the optimal measurement very nearly produces either $\ket{\alpha}$ or $\ket{-\alpha}$ with equal probability, i.e. it collapses the cat. Some alternative choices of binary decision problem which could be thought to reflect a cat size (\textit{e.g.} distinguishing the cat state from the mixed state of the branches, or distinguishing either of the branches from the cat state) do not have this nice property. Quantitatively, applied to $\ket{\psi_{+}}\bra{\psi_{+}}$, the POVM which distinguishes $\ket{\psi_{+}}\bra{\psi_{+}}$ from $0.5\ket{\alpha}\bra{\alpha} + 0.5 \ket{-\alpha}\bra{-\alpha}$ returns $\ket{\psi_{+}}$ with unit probability. For $\vert \alpha \vert^{2} \rightarrow 
\infty$ (and alredy nearly so for $\vert \alpha \vert^{2}\gtrsim 10$), applying the POVM which distinguishes the cat state from $\ket{\alpha}\bra{\alpha}$ to the cat state and subsequently performing a projective measurement in the approximately orthonormal basis $\lbrace \ket{\alpha} , \ket{-\alpha}\rbrace$ gives $\ket{\alpha}$ with probability $1/2 + 1/2\sqrt{2}$ and $\ket{-\alpha}$ with the complementary probability.

\section{\label{sec:wignerhcs2}Wigner function of $\ket{\text{HCS}_{2}(\alpha)}$}

In Section \ref{sec:fisher} we noted that the Wigner function of $\ket{\text{HCS}_{N}(\alpha)}$ (Eq.(\ref{eqn:hierarchical})) does not exhibit the same peak distribution as the Wigner function of $\ket{\Omega}$. Here we explicitly write the Wigner function for $\ket{\text{HCS}_{2}(\alpha)}$: \begin{widetext} \begin{small}

\begin{eqnarray}
W^{\text{HCS}_{2}(\alpha)}(\gamma_{1},\gamma_{2})&:=&{4\over \pi^{2}}\big \langle D_{1}(\gamma_{1})D_{2}(\gamma_{2})e^{i\pi\sum_{i=1}^{2}a^{\dagger}_{i}a_{i}}D_{1}(-\gamma_{1})D_{2}(-\gamma_{2}) \big \rangle_{\ket{\text{HCS}_{2} (\alpha)}} \nonumber \\
&=& {2\over \pi^{2}}\sum_{\epsilon \in \lbrace 1,-1 \rbrace }{1\over (2+2\epsilon e^{-2\vert \alpha \vert^{2}})^{2}} \left( \vphantom{\sum_{\kappa , \tau \in \lbrace 1,-1 \rbrace }} 2\epsilon e^{-2\vert \gamma_{2}\vert^{2}}\cos(2\alpha \text{Im}\gamma_{2})(e^{-2\vert \alpha - \gamma_{1}\vert^{2}}+e^{-2\vert \alpha + \gamma_{1}\vert^{2}}) \right. \nonumber \\ &{}& \left. +   2\epsilon e^{-2\vert \gamma_{1}\vert^{2}}\cos(2\alpha \text{Im}\gamma_{1})(e^{-2\vert \alpha - \gamma_{2}\vert^{2}}+e^{-2\vert \alpha + \gamma_{2}\vert^{2}})+ 4e^{-2(\vert \gamma_{1} \vert^{2}+\vert \gamma_{2}\vert^{2})}\cos(2\alpha \text{Im}\gamma_{2})\cos(2\alpha \text{Im}\gamma_{1}) \right. \nonumber \\ &{}&  \left. + \sum_{\kappa , \tau \in \lbrace 1,-1 \rbrace }e^{-2\vert \alpha +\kappa \gamma_{1} \vert^{2}-2\vert \alpha +\tau \gamma_{2} \vert^{2}} \right) + {2\over \pi^{2}} {1\over 2(1- e^{-4\vert \alpha \vert^{2}})^{2}}\left( e^{-2\vert \alpha - \gamma_{1} \vert^{2}-2\vert \alpha - \gamma_{2} \vert^{2}} + e^{-2\vert \alpha + \gamma_{1} \vert^{2}-2\vert \alpha + \gamma_{2} \vert^{2}} \right. \nonumber \\ &{}& \left. - e^{-2\vert \alpha + \gamma_{1} \vert^{2}-2\vert \alpha - \gamma_{2} \vert^{2}} - e^{-2\vert \alpha - \gamma_{1} \vert^{2}-2\vert \alpha + \gamma_{2} \vert^{2}}  - 4e^{-2(\vert \gamma_{1} \vert^{2}+\vert \gamma_{2}\vert^{2})}\sin(2\alpha \text{Im}\gamma_{1})\sin(2\alpha \text{Im}\gamma_{1})  \right)
\label{eqn:wignerhcs}
\end{eqnarray}
\end{small}
\end{widetext}
with $D_{i}$ the displacement operator of the $i$-th mode, and $P_{\text{tot}}=e^{i\pi\sum_{i=1}^{2}a^{\dagger}_{i}a_{i}}$ the total parity operator. The many-mode Wigner function defined above is the same as that defined by the Fourier transform of the symmetric-order characteristic function.\cite{glauber}

\begin{acknowledgments}
We thank the Kavli Institute for Theoretical Physics program ``Control of Complex Quantum Systems" for hospitality and for supporting this research in part by the National Science Foundation Grant
No. PHY11-25915. This work is also supported by NSF Grant No. CHE-1213141.
\end{acknowledgments}

\bibliography{photon_references.bib}

\end{document}